\pdfoutput=1
\pdfoutput=1
\pdfoutput=1
\pdfoutput=1
\pdfoutput=1
\documentclass[journal=jacsat,manuscript=article]{achemso}
\usepackage{tcolorbox}
\usepackage[version=3]{mhchem} 

\author{Javad Shabanpour}
\email{m.javadshabanpour1372@gmail.com}
\affiliation[Iran University of Science and Technology]
{Department of Electrical Engineering, Iran University of Science and Technology, Narmak, Tehran 16486-13114, Iran}
\author{Sina Beyraghi}
\author{Homayoon Oraizi}
\title[An \textsf{achemso} demo]
{Reconfigurable honeycomb metamaterial absorber having incident angular stability}

\abbreviations{IR,NMR,UV}
\keywords{American Chemical Society, \LaTeX}
\begin{document}

\begin{abstract}
Ultrawide-angle  electromagnetic wave absorbers with excellent mechanical properties are required in many diverse applications such as sensing, and stealth technologies. Here, a novel 3D reconfigurable metamaterial absorber (MMA) consisting of honeycomb and VO2 films is proposed. The proposed MMA exhibits a strong absorptivity above 90\% in the widest incident angle  up to $87^\circ$ for TM- and TE polarized oblique incidences for THz wave propagating in yoz- plane.
According to simulation results, under normal incidence, when VO2 films are in the insulating state, the proposed absorber exhibits high absorptivity in the frequency band of 1-4 THz. By increasing the temperature of the whole structure, the structural transformation of VO2 occurs and turns into the metallic phase. We have shown that under oblique incidence, the ohmic losses of VO2 films especially those parallel to the direction of the  incident electric field are the most important absorption principles of the proposed MMA.
Furthermore, to understand the physical mechanism of absorption, the induced electric field as well as the power loss density of the proposed structure are investigated.
In addition, it is shown that the presented VO2 based honeycomb absorber retains its full-coverage incident angle characteristics for TM-polarized incidences propagating in the xoz-plane.
Due to the ultra wide-angle absorption (angular stability) and mechanical performance, it is expected that the presented MMA may find potential applications, such as camouflage technologies, electromagnetic interference, imaging,  and sensing.
To the best knowledge of authors, the proposed MMA configuration exhibits the absorptivity in the widest incident angle ever reported.
\end{abstract}

\section{Introduction}
Artificial metamaterials composed of subwavelength engineered scatterers have attracted massive attention owing to their abilities to modify the permittivity and permeability values to reach beyond those of composite materials found in nature, which may be structured for complex manipulation of electromagnetic
(EM) waves\cite{1,2,3,4}. The metamaterial absorbers (MMA),  as an interesting application of metamaterials, have become a research hotspot in the past decade and have also been of great interest for solving electromagnetic interference problems, such as the stealth technologies, solar cells and sensor applications\cite{5,6,7,43,44}. In 2008, Landy et al. introduced the first perfect MMA in which electric and magnetic resonances were generated in a narrow frequency band around 11.65 GHz\cite{8}. MMAs have certain advantages compared with conventional
absorbers like ferrite\cite{9} and Salisbury screen absorbers\cite{10}. For instance, MMA can achieve a high level of absorptivity in spite of a thin substrate. Moreover, reconfigurable absorbers can be designed by tunable devices or materials\cite{11,12}. Such great features made MMAs a promising candidate for various applications in the frequency spectrum, from microwaves to
optical signals\cite{13,14}. Since MMAs are made of periodic arrays of resonators, they can only realize efficient absorption in a narrow bandwidth. To solve this problem, several efforts have been made to broaden the absorption frequency band\cite{15,16,17}.

However, in all of the above strategies, the maximum absorption is obtained for the normal incidence and the absorption efficiency degrade for wider incidence angles. In general, the absorption characteristics of MMAs depend on the incident angle and wave polarization. The realization of polarization-independent MMAs is not difficult. It will be feasible for the 
symmetrical unit-cells placed in the vertical and horizontal directions
\cite{18,19}. On the other hand, the design of incident-angle-independent MMAs is a challenging task. Although several works have been attempted to address wide incident-angle MMAs, their maximum coverage of incident angle
 is limited to $70^\circ$\cite{20,21,22}. In 2016, Shen et al., inspired by origami, found that the folded resistive patch array standing up on a metallic plate can exhibit a wide-angle absorbing characteristic up to $75^\circ$ under the transverse magnetic (TM) polarization\cite{23}. But for practical applications, besides possessing outstanding electromagnetic properties, wide-angle MMAs need to be strong enough and have excellent mechanical performance. Therefore, the above 3D absorber has poor mechanical properties which significantly hinders its further usage in real-world practical applications. On the other hand, recent research has revealed that structures with honeycomb cores have excellent mechanical
performances and also can be used for electromagnetic wave
absorption\cite{24,25}. Based on such advantaged, we present a novel vanadium dioxide (VO2)-based honeycomb-like MMA which covers almost the full incident angle ($0^\circ$ to $87^\circ$).  

VO2 is a well-known smart material, which is known and utilized for its ultrafast and brutal reversible phase transition from insulator to metallic state above the critical temperature around ${{\rm{T}}_{\rm{c}}}{\rm{ = 340K}}$\cite{26}. This metal-insulator transition (MIT) due to the atomic level deformation in VO2 can be provoked by thermal\cite{27}, optical\cite{28}, or electrical stimuli\cite{29}. MIT can occur within an order of several nanoseconds or even in picoseconds range for optical stimulation\cite{30}. During MIT,  the electrical conductivity has dramatic changes and can shift up to four orders of magnitude across the two phases\cite{31}. Due to an ultrafast switching time, almost near room critical temperature and useful structural transition characteristics, VO2 has been identified as noteworthy material in reconfigurable metamaterial devices over a broad spectral range from GHz to optics. It has numerous practical applications at terahertz
frequencies such as reprogrammable digital metasurfaces, THz waves modulator, and tunable antennas\cite{32,33}.

In this paper, a novel three-dimensional (3D) MMA is proposed, which consists of hexagonal honeycombs with VO2 thin films deposited on its walls as depicted in \textbf{Fig. 1}. The proposed absorber is engineered in such a way that it can retain the perfect absorptivity in a super-wide incident angle. In comparison to the class of wide-angle absorbers, our proposed MMA provides the widest coverage of incident angles, ever reported to the authors’ knowledge. Since the design of the absorbers working under the oblique incident angles more than $75^\circ$ has not yet been realized and reported due to its complexity, here for the first time, benefiting from VO2 exotic properties and tuning the proper electrical resistivity of VO2, we presented the MMA which can operate up to $87^\circ$ for TM- and TE- polarized oblique incidences for THz waves propagating in a yoz-plane. To demonstrate the main mechanism of absorption, the
induced electric field as well as the power loss density of the proposed Reconfigurable Honeycomb Absorber (RHA) are analyzed. We believe that the 
full coverage of incident-angle characteristics of the proposed RHA  dramatically broadens the range of its applications in various fields such as imaging, sensing, and camouflage technology.
\section{Results}
\textbf{Fig. 1(a)-(b)} shows the front view and top view of the proposed 3D RHA vertically mounted above a gold ground film. The periodicity of the proposed RHA unit-cells is ${D_x} = 68.62\mu m$ and ${D_y} = 40.7\mu m$ along the x and y directions, respectively. 
\begin{figure}[t]
	\centering
	\includegraphics[height=8cm]{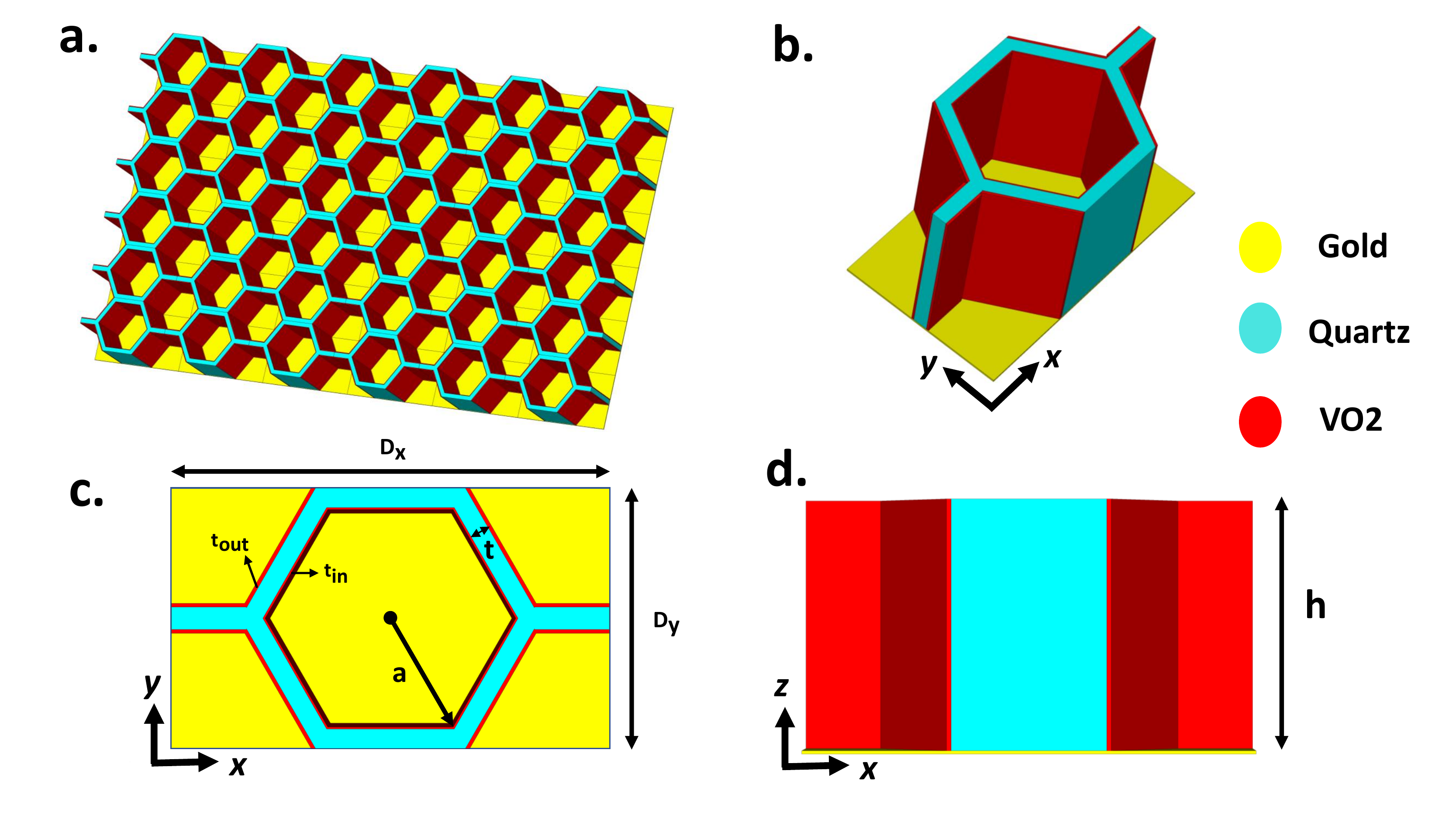}
	\caption{The structural design of the proposed RHA. (a) Perspective view. (b) Unit cell diagram. (c) Front view and (d) side view of the unit cell.}
	\label{fgr:example2col}
\end{figure}
The other geometrical parameters are $h = 38\mu m$, $a = 20\mu m$, $t = 3.5\mu m$, ${t_{in}} = 300nm$ and ${t_{out}} = 600nm$, as shown  in \textbf{Fig.1} respectively. The material of honeycomb is made of quartz with ${\varepsilon _r} = 3.75$ and $\tan \delta  = 0.018$. The inner wall of the honeycomb in each unit of RHA is composed of 6 VO2 films that are joint together and the other 6 VO2 films are deposited on the outer walls of the honeycomb. The thickness of VO2 thin film on the inner and outer walls of the honeycomb are set to be 300nm and 600nm, respectively. The complex dielectric properties of the VO2 thin films can be defined by the Bruggeman effective-medium theory in the THz range, where ${\varepsilon _d}$ and ${\varepsilon _m}$ indicate the dielectric constant of the insulator and metallic regions, respectively and $V$ denotes the volume fraction of metallic regions\cite{34}.
\begin{equation}
 {\varepsilon _{V{O_2}}} = \frac{1}{4}\{ {\varepsilon _d}(2 - 3V) + {\varepsilon _m}(3V - 1) + \sqrt {{{[{\varepsilon _d}(2 - 3V) + {\varepsilon _m}(3V - 1)]}^2} + 8{\varepsilon _m}{\varepsilon _d}} \} \end{equation}
At room temperature, the dielectric constant of VO2 thin film is about 9 in the insulting state\cite{35} and by increasing the temperature of the structure through a resistive heater beneath the gold plate, the structural transformation occurs and VO2 turns into the metallic phase. VO2 films at THz frequencies display electrical conductivity in the range of $10 \sim 100S/m$ in the insulating state\cite{35} and as high as an order of $5 \times {10^5}S/m$ in the metallic state\cite{36,37}. The bottom gold layer with the conductivity of $\sigma  = 4.11 \times {10^7}\,{\rm{S/m}}$ plays an important role as a mirror in impeding the EM waves through the RHA. All the Full-wave EM simulations have been accomplished here by the commercial program CST Microwave Studio. For evaluating the reflection characteristics of the infinite array of RHA meta-atoms, the open boundary condition is applied along the z-axis, whilst periodic boundary conditions are also assigned along the x- and y-directions to incorporate the mutual coupling effect among the   neighboring elements. Meanwhile, TE and TM polarized plane waves with different incidence angles were transmitted to the 3D RHA array along the z-axis.

\subsection{Wave absorbing properties}
For a metamaterial absorber, the absorptivity can be calculated by Eq. (2), where $A(\omega )$, $\Gamma (\omega )$ and $T(\omega )$ are the absorption, reflectance, and transmittance, respectively\cite{38}.
\begin{equation}
A(\omega ) = 1 - \Gamma (\omega ) - T(\omega )
\end{equation}
Therefore, high absorptivity can be obtained by minimizing both the reflection and transmission coefficients. Since the gold plate thickness is much larger than the penetration depth of the incident  wavefronts, $T(\omega )$ equals to zero and the absorptivity of the designed RHA can be simply computed by $A(\omega ) = 1 - \Gamma (\omega )$. Under normal incidence, the reflection coefficient can be expressed as:
\begin{equation}
\Gamma (\omega ) = \frac{{Z(\omega ) - {Z_0}}}{{Z(\omega ) + {Z_0}}}
\end{equation}
where ${Z(\omega )}$ and ${{Z_0}}$
denotes the impedances of the RHA and free space, respectively. Eq. (3) shows that the zero reflection condition is satisfied when ${Z(\omega )}$ and ${{Z_0}}$ are matched.
Within the effective medium approximation, the  impedance of a metamaterial can be controlled by tailoring the effective permittivity ${\varepsilon _r}$ and the permeability ${\mu _r}$ as follows:\cite{39}
\begin{equation}
Z(\omega ) = \sqrt {\frac{{{\mu _0}{\mu _r}(\omega )}}{{{\varepsilon _0}{\varepsilon _r}(\omega )}}} 
\end{equation}
where ${{\varepsilon _0}}$ and ${{\mu _0}}$ are the permittivity and permeability of the free space, respectively. By properly adjusting the VO2 electrical conductivity, the impedance of RHA can be matched to that of the free space. Therefore, the transmission coefficient can be minimized by dissipating the transmitted wave with significant VO2 ohmic losses at the intermediate temperatures. For the lowest and highest orders of electrical conductivity, VO2 is in the dielectric or metallic steady-state phases, respectively. On the other hand, the ohmic losses of the meta-atom are maximum at the intermediate temperatures which lead to a sharp drop in the reflection amplitude, so that the maximum absorption efficiency is attained in the entire frequency range of interest\cite{33}. Therefore, in this condition, the loss factor of the RHA is high because of the large imaginary part of the refractive index (n).

Under the illumination of normal incident wave and considering the  characteristics of VO2, numerical simulations have been performed by choosing the electrical conductivity of $\sigma  = 8000S/m$. The simulated reflection and absorption spectra of the designed RHA for TE and TM polarized incident wave are depicted in \textbf{Fig. 2(a)} and \textbf{Fig. 2(d)}, respectively. The array has a near-unity absorbance in the frequency band of 1-4 THz. Note that for frequencies higher than 4 THz, the maximum periodicity of the structure exceeding one wavelength and the higher-order Floquet modes are generated. Consequently, the proposed structure exhibits a wideband absorptivity above 90\% in 1.2-4 THz, leading to the fractional bandwidth as high as 108\%.
For a better insight, the induced electric field distribution, and the power loss density are demonstrated in \textbf{Fig. 2(b),(c)} and \textbf{Fig. 2(e),(f)} for both x- and y-polarizations at 2.5 THz, respectively. Observe that the VO2 films on the inner and outer walls of the honeycomb have stronger electric field density around them in comparison to the other RHA regions. Note that those VO2 films which are parallel to the incident electric field direction, are more excited. Furthermore, it can be concluded from \textbf{Fig. 2}, that the ohmic losses of VO2 films especially those parallel to the incident wave polarization are the most important absorption principles of the RHA.
If the free-space–absorber impedance matching condition is well satisfied, a great portion of the incident THz wave energy can be effectively absorbed by the proposed 3D RHA.
 The normalized impedance (z)
of the RHA can be calculated by \cite{40}
\begin{equation}
z = \sqrt {\frac{{{{(1 + S_{11}^2)}^2} - S_{21}^2}}{{{{(1 - S_{11}^2)}^2} - S_{21}^2}}} \, = \,\frac{{1 + R}}{{1 - R}}
\end{equation}
\begin{equation}
A = 1 - R = \frac{2}{{z + 1}} = \frac{{2\left[ {{\mathop{\rm Re}\nolimits} (z) + 1} \right]}}{{{{\left[ {{\mathop{\rm Re}\nolimits} (z) + 1} \right]}^2} + {\mathop{\rm Im}\nolimits} {{(z)}^2}}} - i\frac{{2{\mathop{\rm Im}\nolimits} (z)}}{{{{\left[ {{\mathop{\rm Re}\nolimits} (z) + 1} \right]}^2} + {\mathop{\rm Im}\nolimits} {{(z)}^2}}}
\end{equation}
\begin{figure}[t]
	\centering
	\includegraphics[height=8cm]{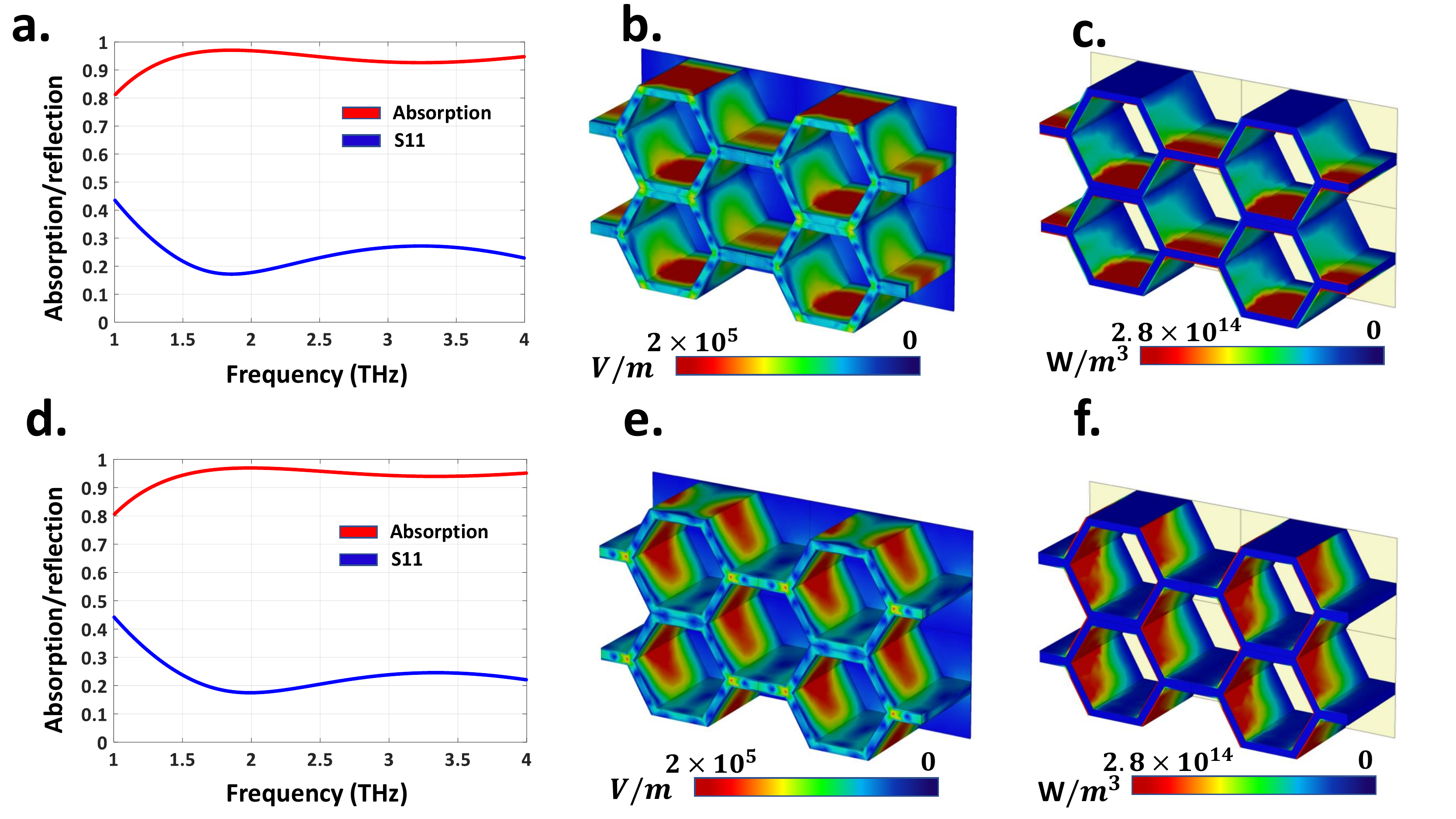}
	\caption{(a), (d) Simulated absorption and reflection spectra of the designed RHA under the illumination of x- and y-polarized normal incidences, respectively. (b), (e) The electric field distributions on the unit cell at 2.5 THz under the illumination of x- and y-polarized normal incidences, respectively. (c), (f) The power loss distributions on the  unit cell at 2.5 THz for x and y normal incidences.}
	\label{fgr:example2col}
\end{figure}
The simulated normalized impedance of the proposed RHA is depicted in \textbf{Fig. 3(a)}. Observe that at resonance frequencies, the real part of the normalized impedance matches the free-space value ${\mathop{\rm Re}\nolimits} (z) \approx 1$, and the imaginary part of the normalized impedance reaches zero simultaneously, ${\mathop{\rm Im}\nolimits} (z) \approx 0$.
Therefore, according to \textbf{Fig. 3(a)}, the principal mechanism of such a wideband absorption behavior of the RHA originates from the multiple resonance characteristics of the proposed structure.

When the incident angle increases, the deterioration of absorptivity is unavoidable since the zero-reflection condition differs under normal and oblique incidences. For example, at oblique incidence, the reflection coefficients for the perpendicular and parallel polarizations can be obtained by:\cite{41}

\begin{equation}
{\Gamma _ \bot }(\omega ) = \frac{{Z(\omega )\cos {\theta _i} - {Z_0}\cos {\theta _t}}}{{Z(\omega )\cos {\theta _i} + {Z_0}\cos {\theta _t}}}
\end{equation}

\begin{equation}
{\Gamma _\parallel }(\omega ) = \frac{{Z(\omega )\cos {\theta _t} - {Z_0}\cos {\theta _i}}}{{Z(\omega )\cos {\theta _t} + {Z_0}\cos {\theta _i}}}
\end{equation}
where ${{\theta _i}}$ and ${{\theta _t}}$ are the incident and transmission angles, respectively. Given that the absorptivity of MMAs changes when the incident angles are varied, so an angle-insensitive unit cell (having angular stability)  must be designed for obtaining a full coverage incident angle MMA.
We will show that our elaborately designed VO2 based unit cell is capable of realizing such a goal.
Increasing the temperature of the whole structure enhances the electrical conductivity of VO2 thin films, and leads to near-unity absorption up to $87^\circ$.
\begin{figure}[t]
	\centering
	\includegraphics[height=8cm]{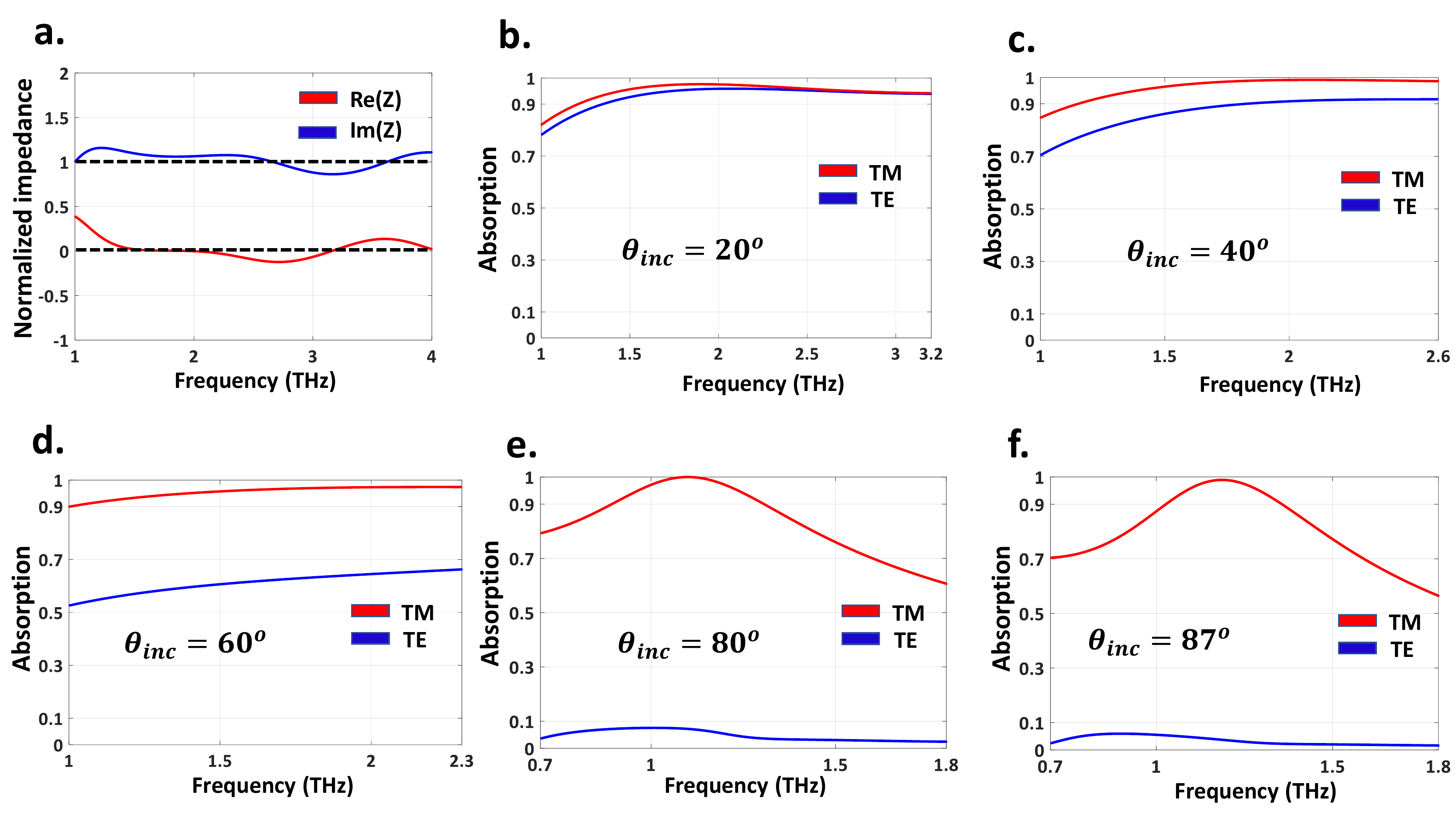}
	\caption{(a) The normalized impedance of the designed RHA versus frequency. The simulated absorption spectra for TE and TM polarization with the different incident angles of (b) $20^\circ$. (c) $40^\circ$. (d) $60^\circ$. (e) $80^\circ$ and (f) $87^\circ$.}
	\label{fgr:example2col}
\end{figure}
\textbf{Fig. 3(b)-(f)} demonstrates the absorption spectra of the RHA under different incident wave angles of TE- and TM-polarized waves propagating in the xoz- plane. Observe that benefiting from the structural transition of VO2 by increasing its temperature, our elaborately designed RHA can retain the absorptivity (by more than 90\%) in a super-wide incident angle up to $87^\circ$ for the TM-polarized wave.
For the incident angles lower than $60^\circ$, VO2 should be in the insulator state ($\sigma  = 8000S/m$), and for the incident angles greater than $60^\circ$, the structural transition in VO2 must occur to metal state ($\sigma  =5 \times {10^5}S/m$).

According to \textbf{Fig. 3}, by increasing the incident angle, the absorbing bandwidth becomes narrow and the full coverage of incident angle of the proposed structure occurs at 1-1.3 THz.
Besides, by increasing the temperature of the structure, the absorptivity of the TE-polarized oblique incidence drops sharply.  
To intuitively understand and interpret the absorption mechanism, the power loss density distributions of the proposed array at various angles are also depicted in \textbf{Fig. 4}. Observe that the power loss can be mostly attributed to the ohmic losses provided by VO2 thin films based on
\begin{equation}
{P_{loss}=\int {\sigma \left| E \right|} ^2}dv
\end{equation}
whereby $E$ is the tangent electric field. \textbf{Fig. 4} shows that the ohmic losses of VO2 films, especially those parallel to the direction of the incident electric field are the most effective factors of absorption of the proposed RHA.
\begin{figure}
	\centering
	\includegraphics[height=8cm]{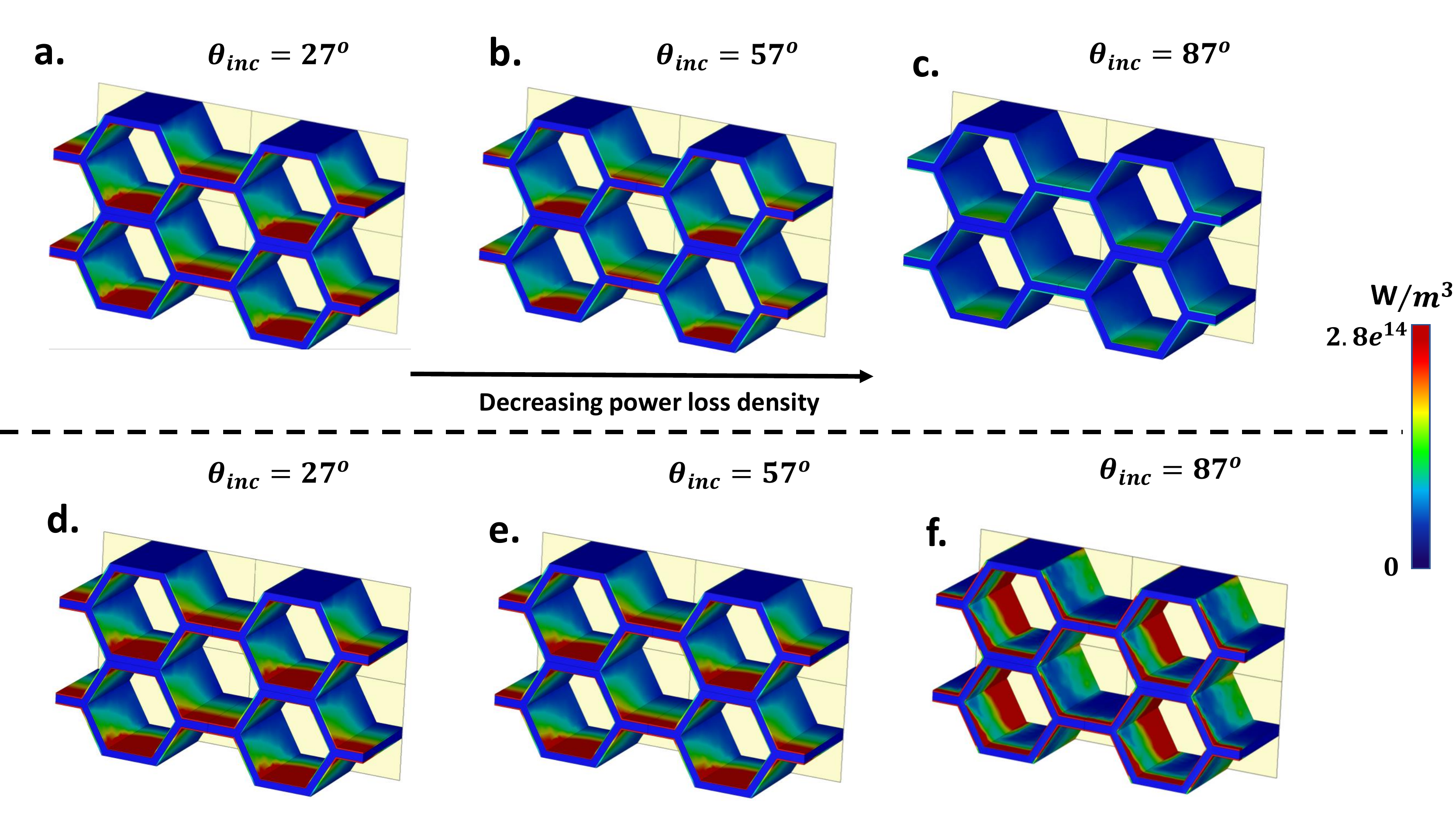}
	\caption{The power loss density distributions on the array at the various TM polarized incident angles of (a). $27^\circ$. (b) $57^\circ$ and (c) $57^\circ$ for electrical conductivity of 8000S/m at 1.3 THz. The power loss density distributions for electrical conductivity of VO2 (d) 8000S/m. (e) 12000S/m and (f) $5 \times {10^5}S/m$ respectively. }
	\label{fgr:example2col}
\end{figure}
\newpage
Observe also from \textbf{Fig. 4(a)-(c)} that for a constant electrical conductivity, with the increase of oblique incident angle, the power loss density decreases. Eq. (9) shows the only way to increase the power loss density is to increase the electrical conductivity of the films.  
As depicted in \textbf{Fig. 4(d)-(f)}, increasing the electrical conductivity of VO2 which can be dynamically tuned, plays an effective role in dissipating the incoming EM energy.
Furthermore, when the oblique incident angle increases, the power loss distributes on the lateral sides of VO2 films instead of the upper and lower films.

\textbf{Fig. 5} shows the absorption spectra of the proposed RHA under different incident wave angles of TE- and TM-polarized waves in both xoz- and yoz-planes.
For TM-polarized wave, as mentioned earlier, the near-unity absorption can be obtained up to $87^\circ$ for the oblique incident wave propagating in the xoz- plane.  Observe in \textbf{Fig. 5(b)} that this behavior also happens for THz waves propagating in the yoz- plane.
We have mentioned in \textbf{Fig. 3} that by increasing the temperature of the structure, the absorptivity of the TE-polarized oblique incidence drops sharply. But for different values of electrical conductivity of VO2 (See \textbf{Fig. 5(c)} ),  our elaborately designed structure can retain the absorptivity up to $87^\circ$  for an oblique incident wave propagating in the yoz- plane. While for TE-polarized oblique incidence propagating in the xoz- plane, the effective permittivity of the RHA will gradually drop as long as the angle of incidence increases; therefore, the absorption peaks shift towards the higher frequencies (higher than 2.3 THz) where the higher-order Floquet modes are generated as can be observed from \textbf{Fig. 5(d)}. On the contrary, for the TM polarized wave, since the magnetic component of the incident wave is always perpendicular to the incidence plane at different incident angles, the anti-parallel currents are effectively excited, leading to an ultra-wide-angle absorption for TM-polarized oblique incidences for THz wave propagating in both xoz- and yoz- planes\cite{42}.

\begin{figure}[t]
	\centering
	\includegraphics[height=8cm]{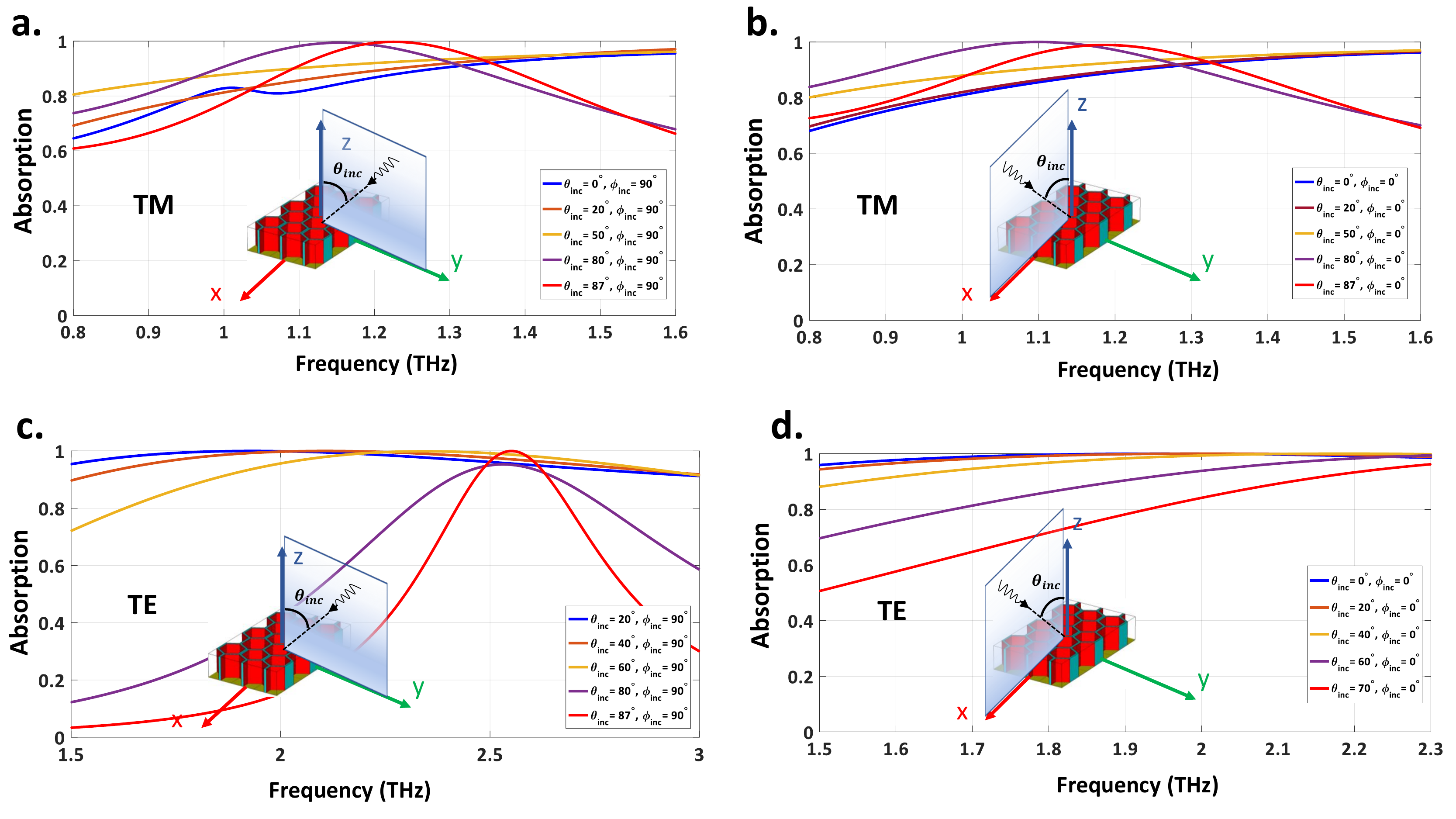}
	\caption{The simulated absorption spectra for four different cases of: (a) TM-polarized incidence in yoz-plane. (b) TM-polarized incidence in xoz plane. (c) TE-polarized incidence in yoz-plane and (d) TE-polarized incidence in xoz-plane with the different incident angles.}
	\label{fgr:example2col}
\end{figure}
\section{Discussion}
In summary, we designed a novel reconfigurable metamaterial absorber that can support a good absorption for an almost complete incident angle range (having angular stability) by changing the electrical conductivity of the VO2 thin films placed onto the inner and outer walls of the honeycomb.
For TM- and TE polarized oblique incidences for the THz wave propagating in the yoz- plane, the proposed RHA exhibits a strong absorptivity above 90\% up to the incidence angle of  $87^\circ$. Besides, the proposed configuration of the RHA consists of hexagonal honeycombs cores which have excellent mechanical performances. The ultra-wide-incident angle property (angular stability) of the RHA was justified through analyzing the induced electric field as well as the power loss density distributions.
We have demonstrated that, by increasing the electrical conductivity of VO2 which can be dynamically tuned, the ohmic losses of VO2 films especially those parallel to the direction of  incident electric field are the most effective absorption factors of the proposed RHA. Furthermore, as the oblique incident angle increases, the power loss tend to distribute on the lateral sides of VO2 films instead of the upper and lower films. We believe that the proposed absorber may find great potential for engineering applications due to its angular stability and full incident angle absorption and mechanical performance.
\section{Competing interests}
The author declare no competing interests.




\end{document}